\documentstyle[multicol,prb,aps,epsf] {revtex}
\begin{document}
\bibliographystyle{prsty}
\draft
\title{Relaxation of frustration and gap enhancement 
        by the lattice distortion in the $\Delta$ chain}
\author{Tota Nakamura $^1$ and Satoshi Takada $^2$}
\address{$^1$ Department of Applied Physics, Tohoku University,
         Sendai, Miyagi 980-77, Japan\\
         $^2$ Institute of Physics, University of Tsukuba,
         Tsukuba, Ibaraki 305, Japan }
\date{\today}
\maketitle
\begin {abstract}
We clarify an instability of the ground state of the $\Delta$ chain 
against the lattice distortion that increases a strength $(\lambda)$ 
of a bond in each triangle.
It relaxes the frustration and causes a remarkable gap enhancement; 
only a $6\%$ increase of $\lambda$ causes the gap doubled from 
the fully-frustrated case $(\lambda=1)$.
The lowest excitation is revealed to be a kink-antikink bound state
whose correlation length decreases drastically with $\lambda$ increase.
The enhancement follows a power law,
$\Delta E_{\rm gap}\sim (\lambda-1) + 1.44 (\lambda -1)^{\frac{2}{3}}$,
which can be obtained from the exact result of the continuous model.
This model describes a spin gap behavior of the delafossite YCuO$_{2.5}$.
\end  {abstract}

\begin{multicols}{2}

\narrowtext

Cooperations between theory and experiment in the field of the low-dimensional
quantum spin systems are of greater importance especially in these years,
since various new compounds have been synthesized which realize
theoretical models. \cite{ramirez94}
Much interest recently is focused especially on the system with a spin gap.
As typical examples, we can consider
the spin ladder model,\cite{ladders} the bond-alternation model,\cite{alters} 
and the Majumdar-Ghosh model.\cite{majumdar-g69}

The ground state of the $\Delta$ chain 
shares common properties with models stated above. 
Its ground state is the pure singlet-dimer state with two-folded degeneracies.
The existence of the excitation energy gap above the ground state is
rigorously proven.\cite{monti-s91,monti-s92}
However,
strong frustration in this model plays an important role to cause
unusual properties in the excited states, i.e.,
its excitation spectrum is dispersionless (zero-energy mode).\cite{kubok93}
This mode contributes to the existence of the additional low-temperature
peak in the specific heat.\cite{nakamura-s95,otsuka95}
One of the authors (T.N.) and Kubo,\cite{nakamura-k96} 
and Sen et al \cite{sen-swc96}
clarified its elementary excitations and how they contribute 
to the thermodynamic properties.
They found that the properties of the excitation is 
governed by a kink and an antikink.
The dispersionless aspect originates in a localized kink, 
while an antikink moves as a free particle with an effective 
mass between two localized kinks.
This picture is in a clear contrast with the excitation of the Majumdar-Ghosh 
model, where both a kink and an antikink are 
equivalently mobile.\cite{shastry-s81}
Sen et al \cite{sen-swc96} also pointed out the $\Delta$ chain 
can be a model for the recently discovered delafossite YCuO$_{2.5}$,
but the estimated spin gap is about half the one measured in the experiment.

\begin{figure}[h]
\epsfxsize = 8cm
\epsffile{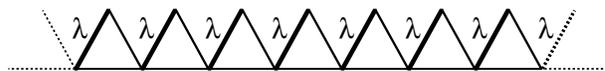}
 \caption {Shape of the distorted $\Delta$-chain.
           Double lines indicate the $\lambda$ bonds.
  \label{fig:lattice}
          }
\end  {figure}

Strongly frustrated quantum spin systems have been attracting, 
since the interplay of the quantum effects and frustration 
may lead to an exotic ground state.
Our question in this letter is that such a spin gap state caused by 
strong frustration is whether stable or not against a perturbation.
If relaxation of frustration occurs by a small perturbation, 
it may change the spin gap.
A pairwise dimerization on the $\lambda$ bonds shown in Fig. \ref{fig:lattice}
can be realized by the lattice distortion in this system,
if the merit of the gap enhancement overcomes the distortion cost.
Therefore we investigate the pairwise-dimerized 
$\Delta$ chain in this letter and clarify its gap behavior.
The analyses are based upon
numerical diagonalization, the variation and the continuous limit. 
First, we discuss the ground state and find a parameter region 
that can be realized in the material, and then we
study the excitation.
In due course,
we will find a remarkable enhancement of the gap due to relaxation of 
frustration by a small increase of the pairwise bond strength ($\lambda$);
only a $6\%$ increase of $\lambda$ causes the spin gap doubled,
which is consistent with the gap measured in YCuO$_{2.5}$.

We consider the following Hamiltonian as a model dimerized by lattice 
distortion.
\begin{equation}
 H=\sum_{n=1}^N h_n
\end  {equation}
with
\begin{equation}
 h_n= \lambda\mbox{\boldmath $S$}_{2n-1}\cdot\mbox{\boldmath $S$}_{2n  }
           + \mbox{\boldmath $S$}_{2n  }\cdot\mbox{\boldmath $S$}_{2n+1}
           + \mbox{\boldmath $S$}_{2n-1}\cdot\mbox{\boldmath $S$}_{2n+1}
\end  {equation}
Here, $N$ is the number of the triangles in the system, $\lambda$ is 
a parameter denoting the dimerization, and $|\mbox{\boldmath $S$}|=1/2$.
Figure \ref{fig:lattice} shows the depicted lattice.
In this letter,
we mainly consider the parameter $\lambda$ near the symmetric point 
$\lambda=1$.

\begin{figure}[h]
\epsfxsize = 8cm
\epsffile{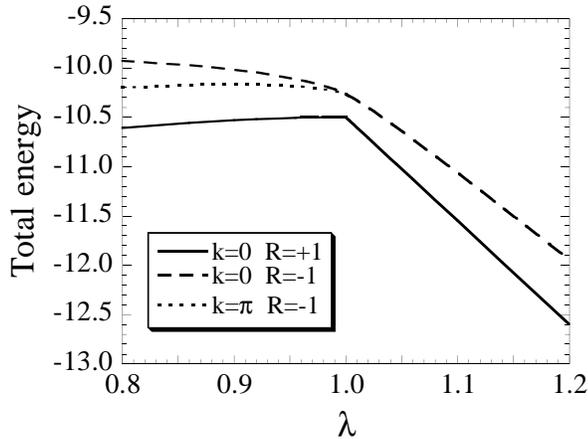}
 \caption{The $\lambda$ dependence of the energy of the
          lowest state in each subspace denoted by $k$ (wave number), 
          and $R$ (spin reversal symmetry) in the system with $N=14$ 
          (28 spins).
 \label{fig:diagE} 
         }
\end  {figure}

We numerically diagonalized the above Hamiltonian up to the systems with 
28 spins $(N=14)$ under the periodic boundary conditions. 
The $\lambda$ dependence of the energy is shown in Fig. \ref{fig:diagE}.
The phase space is divided into a subspace characterized by 
eigenvalues of the wave number ($k$) and the spin reversal symmetry ($R$)
in order to reduce the memory use.
From a technical reason, we only diagonalized the space with 
the wave number $k=0$ and 
$k=\pi$, where the ground state and the relevant low-lying excited states
are found when $\lambda=1$.
It is noticed that the ground state at the fully-frustrated point, 
$\lambda=1$, is most unstable.

The ground state in the case of $\lambda > 1$ is trivial, i.e.,
the ground of the local Hamiltonian $h_n$ is the singlet dimer 
state on the $\lambda$ bond.
Direct product of the dimers can span the whole system and be the 
exact ground state.
Thus the total ground state energy is $-0.75\lambda N$.
Two excited states shown in the figure degenerate for $\lambda > 1$, 
which was also observed for higher states.
This degeneracy suggests that the excitation is local.

On the other hand, the ground state for $\lambda < 1$ is not trivially solved.
It lies in the space of $k=0$, and the ground state energy 
gradually decrease with $\lambda$ go away from 1.
A variational analysis and the second-order perturbation analysis 
manifest that the energy difference begins with the order $(1-\lambda)^2$.
(Details are reported elsewhere.)
Therefore the ground state is more stabilized by the perturbation with 
$\lambda > 1$, which may be actually realized.
In this letter, 
We only discuss the case $\lambda > 1$ in detail, and leave the 
case $\lambda < 1$ for another opportunity.

\begin{figure}
\epsfxsize = 8cm
\epsffile{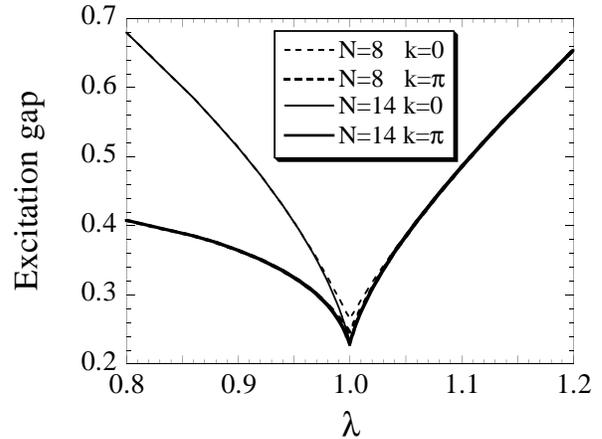}
\caption{$\lambda$ dependence of the first excitation gap in $k=0$ and 
         $k=\pi$ sector.
         Sizes of the systems are $N=8$ and $N=14$.
         For $\lambda < 1$, 
         gap in $k=\pi$ becomes smaller than that in $k=0$.
\label{fig:gaps}
        }
\end  {figure}

Figure \ref{fig:gaps} shows the $\lambda$ dependence of the first 
excitation gap for both $k=0$ and $k=\pi$ sector.
Data of sizes $N=8$ and $N=14$ are compared to see the size dependence,
as well.
Degeneracy between ``$k=0$'' and ``$k=\pi$'' 
is solved for $\lambda < 1$, and ``$k=\pi$'' becomes the first excitation.
The gap enhancement is remarkably large for $\lambda > 1$.
Size dependence is also very weak for $\lambda$ away from 1.
At $\lambda=1.06$, the gap is about twice the one at $\lambda=1$.

We consider the following open system to reveal the excited state.
\begin{equation}
 H_{\rm open}=\sum_{n=1}^{N} h_n  
    + \lambda\mbox{\boldmath $S$}_{2N+1}\cdot\mbox{\boldmath $S$}_{2N+2}.
\end  {equation}
We estimate the excitation energy by the following variational analysis.
The excited state we consider as the first approximation 
consists of $(N-1)$ singlet dimers and two free spins. 
As was done in the previous paper,\cite{nakamura-k96} we name the free spins
a kink and an antikink.
The wave function of an antikink for an improved approximation may 
span beyond several spins since the free spin is not
an eigenstate of the Hamiltonian.
We have considered an antikink consists of 5 spins previously, but
it merely increase the effective mass of an antikink and 
the shape of the wave function are not 
significantly influenced.\cite{nakamura-k96,sen-swc96} 

\begin{figure}[ht]
\epsfxsize = 8cm
\epsffile{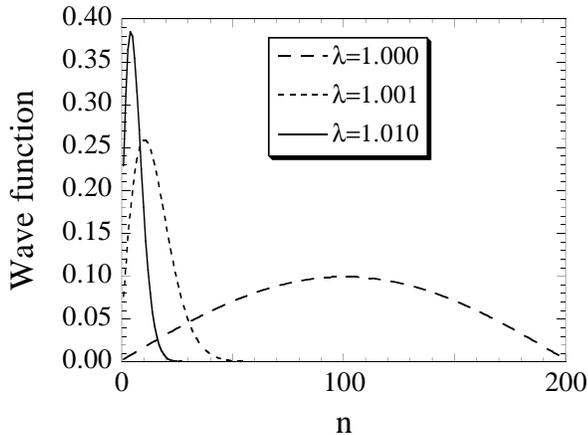}
 \caption {Variational wave function of the lowest eigenstate 
           for $\lambda=1.00, 1.001$, and $1.01$.
           Size of the system $N=200$.
           $n$ stands for the site number of the triangles.
 \label{fig:wavef}
          }
\end  {figure}

We define a 
variational basis $\psi_i$ so that an antikink is located at the $i$th
triangle.
\begin{eqnarray}
 \psi_i \equiv  \psi_{\rm kink}&\otimes& [4,5] \cdots [2i-2,2i-1] \alpha_{2i} 
                \nonumber \\
       &      &         [2i+1, 2i+2] \cdots [2N+1, 2N+2],
\end  {eqnarray}
where $[i, j]$ denotes a singlet dimer state connecting the $i$th and 
the $j$th site,
namely $[i, j]=\alpha_i\beta_j-\beta_i\alpha_j$ for
$\alpha_i (\beta_i)$ denoting up (down) spin located at the $i$th site.
$\alpha_{2i}$ is an antikink.
The wave function of a kink located at the leftmost edge is known as,
\begin{equation}
 \psi_{\rm kink}=[\alpha_1(\alpha_2\beta_3-\beta_2\alpha_3)
                 +\alpha_2(\alpha_1\beta_3-\beta_1\alpha_3)]/\sqrt{6}.
\end  {equation}
This state is an eigenstate of the local Hamiltonian $h_1$, and therefore
does not move.
Its energy eigenvalue is $\lambda /4-1$, therefore a kink contributes to 
the excitation by $(\lambda -1)$.
The singlet dimers that are not on the $\lambda$-bonds, existing between 
a kink and an antikink, are not eigenstates of the local Hamiltonians, $h_n$.
They also contribute to the excited energy.
Variational bases are not orthogonal each other and satisfy the following 
relations.
\begin{eqnarray}
 \langle \psi_i  |\psi_j\rangle&=&\left (-\frac{1}{2}\right )^{|i-j|} 
                               \equiv S_{ij} \label{eq:S}\\
 \langle \psi_i|H|\psi_j\rangle&=&\left[E_{\rm g}
                                 +(\lambda-1) 
                                 +\frac{3}{4}(\lambda-1)\min (i,j) \right]
                                  \langle \psi_i|\psi_j\rangle \nonumber \\
                               & &+\frac{3}{4}\delta_{ij}
                               \equiv H{ij}
\end  {eqnarray}
Here, $\delta_{ij}$ is the Kronecker delta, and $\min (i,j)=i$ if $i\leq j$.
The ground state energy $E_{\rm g}=-0.75\lambda (N+1)$.
Since the matrix $\hat{S}$ given by eq. (\ref{eq:S}) is the positive 
hermitian matrix,
we can easily transform this variational problem with the trial function
$\Psi_{\rm var}\equiv \sum_{i} C_i\psi_i$  into the eigenvalue problem of 
the following Hamiltonian.
\begin{equation}
\hat{H}=(E_{\rm g}+\lambda-1)+\frac{3}{4}\hat{S}^{-1}+\hat{V}
\label{eq:vare}
\end  {equation}
with
\begin{equation}
\hat{V}=\frac{3}{4}(\lambda-1)\hat{S}^{-1/2}\hat{M}\hat{S}^{-1/2},
\end  {equation}
where $M_{ij}=\min (i,j)S_{ij}$.
This can be solved numerically for arbitrary $N$, 
say we show the result of $N=200$ in this letter.
Fig. \ref{fig:wavef} shows the wave function of the lowest eigenstate
for $\lambda=1.00, 1.001, 1.01$.

We find from this figure that an antikink drastically approaches a kink
as $\lambda$ increase from $1$. 

\begin{figure}[ht]
\epsfxsize = 8cm
\epsffile{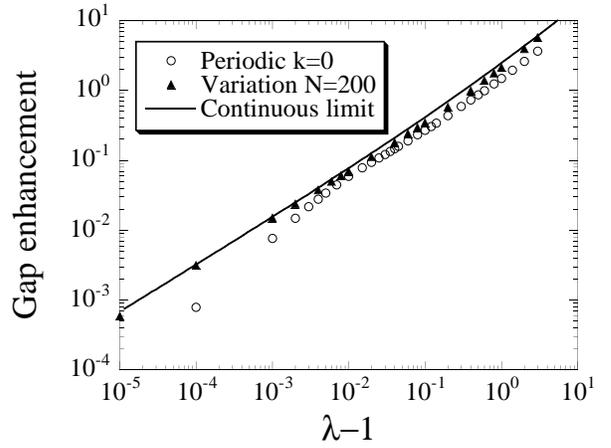}
 \caption {Log-log plot of the gap enhancement for the variational results 
           of $N=200$, and the numerical ones after extrapolated 
           $N\to\infty$. We also plotted the exact result in the continuous 
           limit of eq. ({\protect {\ref{eq:Econt}}}) with the mass 
           $m=1.21$.
 \label{fig:logloggap}
          }
\end  {figure}

Fig. \ref{fig:logloggap} shows
the behavior of the gap enhancement defined by
$ E_{\rm gap}(\lambda)- E_{\rm gap}(1)$.
We also plot the gap enhancement of the periodic system after the extrapolation of 
$N\to\infty$ by using data with $N= 8, \cdots, 14$.
Since the variational wave function of $\lambda=1.01$ only takes a 
finite value for $n < 20$, as shown in Fig. \ref{fig:wavef},
the finite size treated in the numerical diagonalization 
is enough to extract the properties of infinite-size system for 
$\lambda > 1.01$.
Contrary to this, data for $\lambda < 1.01$ deviate from the variational ones,
because the length of the bound state exceeds the finite size.
The gap behavior qualitatively agrees with that of the periodic system,
and is found to obey 
\begin{equation}
 E_{\rm gap}(\lambda)-E_{\rm gap}(1)\simeq
     (\lambda-1)+1.44\times (\lambda -1)^{2/3}.
\end  {equation}
This behavior is what is expected by the following analysis 
in the continuous limit of the model. 

We start by rewriting the variational bases and the representation of the 
Hamiltonian given by eq. (\ref{eq:vare}) into the momentum space.
Since the wave length is defined to be sufficiently larger than the lattice 
constant in the limit, we consider the case of wave number $k\to 0$.
We take into account up to the $k^2$ order for the diagonal part, and 
the leading order for the off-diagonal part.
Then we obtain eigenvalue equations of the following Hamiltonian.
\begin{equation}
 H(\lambda)=\epsilon_0 + (\lambda -1)
-\frac{1}{2m}\frac{d^2}{dx^2}+\frac{3}{4}(\lambda -1)x 
\label{eq:continuous}
\end  {equation}
Here, $x$ is a distance between a kink and an antikink, $m$ is an effective 
mass for an antikink, the constant term 
$\epsilon_0 = E_0(\lambda)+E_{\rm gap}(1)$ 
for $E_0(\lambda)$ is the ground state energy.
The third term of eq. (\ref{eq:continuous}) is a kinetic energy of an 
antikink, and the fourth term is a triangular potential 
expressing the energy loss by 
the singlet dimers between a kink and an antikink.
Within the first approximation that an antikink is a free spin, 
the mass is 1 ($m=1$).
In the exact continuous model, on the other hand,
structure of an antikink only renormalizes the mass as $m(\lambda)$,
as far as the size of an antikink is negligible compared with the 
length of the bound state.

The eigenvalue equation $H\Psi=E\Psi$ 
always gives bound states irrespective of $\lambda$.
This means that a kink and an antikink always form a bound state for 
$\lambda > 1$.
They collapse to a triplet state on a $\lambda$ bond in the limit 
$\lambda\to\infty$.
The wave function of the ground state of eq. (\ref{eq:continuous}) 
is given by the Airy function with the eigenvalue $E$:
\begin{equation}
 E-\epsilon_0= (\lambda -1) 
  +\frac{1}{2}\left(\frac{9}{4m}\right)^{1/3}(\lambda -1)^{2/3}\tilde E
\label{eq:Econt}
\end  {equation}
with $\tilde E=2.33816$.\cite{stern72}
We also plot this expression in Fig. \ref{fig:logloggap} with the mass $m=1.21$,
the value estimated at $\lambda=1$,\cite{nakamura-k96}
though there is not a clear difference with the estimates using $m=1$.
This line coincide with the variational results for $\lambda$ close 
to $1$, but it shifts upward as $\lambda$ increases.
It suggests that the renormalized mass $m(\lambda) > m(\lambda=1)=1.21$.
An averaged length of the bound state is proportional to $(\lambda-1)^{-1/3}$.

\begin{figure}[ht]
\epsfxsize=8.0cm
\epsffile{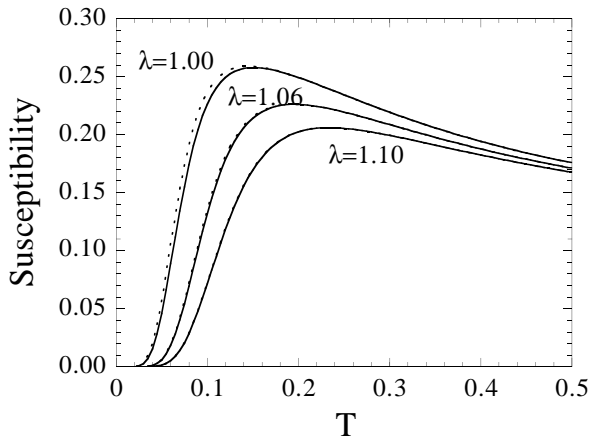}
 \caption{Uniform susceptibility per spin 
          calculated for $\lambda=1, 1.06$ and $1.10$
          by the exact diagonalization of the finite systems with the 
          periodic boundary conditions.
          The sizes of the systems are $N=6$(the dashed line) 
          and $7$(the continuous line).
          As $\lambda$ increase, size dependences become weak.
 \label{fig:suscep}
         }
\end  {figure}

We also calculated the magnetic susceptibility of finite systems 
($N=6$ and $7$) with periodic boundary conditions.
All the eigenvalues were calculated by the numerical diagonalization.
Therefore the numerically exact values were obtained.
Figure \ref{fig:suscep} shows the temperature dependence of the 
susceptibility for $\lambda=1.00, 1.06, 1.10$.

Peak position shifts toward the high temperature side, which is caused 
by the enhanced spin gap with increasing $\lambda$.
We consider the present data is enough to explain the thermodynamic limit 
for $\lambda=1.06$ and $1.10$, since the length of the bound state is
within the finite size.

In summary, 
we have investigated the $\Delta$ chain with the dimerized 
$\lambda$-bond strength increased by a lattice distortion.
For the $\lambda > 1$, the gap was found to obey
$(\lambda-1)+1.44 (\lambda-1)^{2/3}$; 
the first term is contributed by a kink and the latter is by an antikink.
Only $6\%$ change of the bond strength may cause the gap doubled.
Therefore care must be paid for the comparison with the experimental results.
Analyses on the case of $\lambda < 1$ and the detailed derivation of the 
continuous limit will be reported elsewhere.

Recently, Chitra et al investigated the 1-dimensional $J_1$-$J_2$ model with 
the bond alternation, and found that the gap behaves with $\delta^{2/3}$ with 
the alternation parameter $\delta$.\cite{shastry95}
We consider it is the same physics with the present model that 
a kink-antikink bound state is formed by the potential proportional to 
the $\delta$ times distance between a kink and an antikink,
which leads to the gap with exponent $2/3$.

Authors would like to thank Professor Kenn Kubo and Professor 
Atsushi Oshiyama for valuable discussions.
They also acknowledge thanks to Professor Hidetoshi Nishimori for his 
diagonalization package, Titpack Ver. 2.
Computations were performed on Facom VPP500 at the ISSP, University of Tokyo.

\end{multicols}
\end{document}